# ESRF-type lattice design and optimization for the High Energy Photon Source[*]

XU Gang(徐刚)   JIAO Yi(焦毅)   PENG Yuemei(彭月梅)

Key Laboratory of Particle Acceleration Physics and Technology, Institute of High Energy Physics,

Chinese Academy of Sciences, Beijing 100049, China

**Abstract:** A new generation of storage ring-based light source, called diffraction-limited storage ring (DLSR), with emittance approaching the diffraction limit for multi-keV photons by means of multi-bend achromat lattice, has attracted worldwide and extensive studies. Among various DLSR proposals, the hybrid multi-bend achromat concept developed at ESRF predicts an effective way of minimizing the emittance and meanwhile keeping the required sextupole strengths to an achievable level. For the High Energy Photon Source planned to be built in Beijing, an ESRF-type lattice design consisting of 48 hybrid seven-bend achromats is proposed to reach emittance as low as 60 pm·rad with a circumference of about 1296 m. Sufficient dynamic aperture promising vertical on-axis injection and moderate momentum acceptance are achieved simultaneously for a promising ring performance.

## 1 Introduction

Along with the continuous advance in accelerator technology and unceasing pursuing of higher quality photon flux, worldwide attention and efforts are paid to push the brightness and coherence beyond the existing third generation light sources, by reducing significantly the emittance to approach the diffraction-limit for the range of x-ray wavelengths of interest for scientific community. Such new-generation rings are usually called as diffraction-limit storage rings (DLSRs). An international overview of the DLSR design and plans can be seen in Ref. [1]. To achieve an ultralow emittance (e.g., several tens of pm·rads) with the least possible cost, multi-bend achromats (MBAs) with a combination of compact magnets and strong focusing quadrupoles are generally adopted in DLSR designs.

It is the MAX-IV [2] which first adopted small-aperture, high-gradient magnets (e.g., quadrupole gradient of up to 50 T/m with bore radius of 12.5 mm) and small-dimension NEG-coated vacuum chambers in its 7BA design. With these advanced technologies, it was able to realize an emittance of about 300 pm·rad within a circumference of 528 m for a 3 GeV beam. Following MAX-IV, MBA lattice proposals have been raised for a number of facilities being designed, constructed, or upgraded, including Sirius [3], PEP-X [4], ESRF-U [5], APS-U [6], ALS-II [7], Spring-8-II [8] and the High Energy Photon Source (HEPS, originally named BAPS [9-10]). In these proposals, many measures were taken to decrease the emittance to below 100 pm·rad and meanwhile, to keep the beam dynamics robust and satisfactory. For instances, vertical focusing [11] or longitudinal

---

[*] Supported by NSFC (11475202, 11405187) and Youth Innovation Association of Chinese Academy of Sciences (2015009)

gradient [12] is combined into the dipole to shorten the unit cell length and to help minimize the emittance; a high-field (2-3 T) super bend [3] or 3-pole wiggler [5] is inserted at (or close to) the center of the MBA as a hard X-ray source; phase optimization is performed to improve the nonlinear beam dynamics, either by constructing a third-order geometric achromat with several identical MBAs [4] or by employing $-I$ transport between sextupole pairs in a single MBA [5]; quadrupole gradient is further enhanced to approach 100 T/m [5] by using high-permeability pole material or permanent magnet material near the poles to reduced saturation, for a more compact magnet design and MBA layout; moreover, as illustrated in the so-called hybrid-MBA concept [5] first proposed at ESRF, a dispersion bump is created between the outer two dipoles for a more efficient chromatic correction than available in a normal MBA.

HEPS is a kilometer-scale, 5-6 GeV, ultralow-emittance storage ring-based light source, planned to be built in Beijing. Various lattice designs and relevant studies have been performed for HEPS since 2010 [9-10, 13-20]. Since HEPS is a new machine, it has great flexibility in the choice of the ring parameters, e.g., the circumference. It is known that with a larger circumference it will be easier to achieve an ultralow emittance as well as a satisfactory beam dynamics. However, it is necessary to reduce the circumference as much as possible so as to reach a relatively low budget. As a compromise and based on other considerations (e.g., to achieve a harmonic number of 2160 for ~500 MHz RF system and to keep the capability of storing beam of 6 GeV), recently the circumference is determined to be around 1296 m and the lattice cell structure is chosen to be 7BA instead of double-bend or triple-bend achromat.

A PEPX-type lattice [10] has been proposed for the HEPS by using 44 normal 7BAs, with a circumference of 1294.2 m and with emittance of 90 pm·rad for a 6 GeV beam. With delicate optimization, a dynamic aperture (DA) larger than physical aperture and promising off-axis pulsed-sextupole injection, and a momentum acceptance (MA) as large as 3% for a long enough lifetime can be obtained in two separate modes, which are only different in sextupole and octupole strengths and thus can be easily switched from one to the other. Nevertheless, it is very difficult to further push down the emittance in this type of design; otherwise impractically high-gradient or thick sextupoles will be required to correct the increasing natural chromaticities.

As will be shown in Sec. 2, with an ESRF-type lattice design consisting of 48 hybrid-7BAs, a lower emittance of 60 pm·rad can be reached with a similar circumference. In addition, the difficulty of the chromatic correction existed in the PEPX-type design can be greatly mitigated with the dispersion bump. The nonlinear beam dynamics is studied in Sec. 3. It appears feasible to obtain sufficient DA and MA for vertical on-axis injection and a moderate lifetime simultaneously. Conclusions are given in Sec. 4.

**2 Linear optics design**

In the design of a hybrid-7BA, several key demands should be satisfied. First, for the central three unit cells, quadrupoles with strong horizontal focusing and dipoles combined with vertical focusing gradients are required to minimize the emittance and the cell length. Secondly, it needs to create two symmetric dispersion bumps in the gaps between outer dipoles (with as large dispersion as possible between the first and the second, and between the sixth and the seventh dipoles) where sextupoles are installed to correct the natural chromaticity. The third, the phase advance between each pair of sextupoles in a hybrid-7BA should be at or close to $\pi$, thus eliminating most of the undesirable effects of sextupoles. And the fourth, it needs to introduce longitudinal gradient into the outer dipoles (with stronger bending field at the part with greater distance from the dispersion bump), to increase the dispersion at sextupole and to further decrease the emittance.

Based on the above, the hybrid-7BA for HEPS is designed in two steps. First, the case without longitudinal gradient combined in the outer dipoles is considered. The linear optics is matched such that the first three demands mentioned above are satisfied. To make a practical design, as many constraints on magnets and drift spaces as possible are included in the optics matching. For instances, it is required that the maximum focusing gradient is 80 T/m for the quadrupoles in the central three unit cells, and is 50 T/m for others; for the central three combined-function dipoles, the bending radii should be larger than 40 m and the gradients should be smaller than 48 T/m; the length of the long straight section for insertion device (ID) or injection is fixed to 6 m; enough drift spaces are preserved for sextupoles, octupoles, diagnostics, correctors, and for fast feedback kickers (drift between the first and the second quadrupoles and that on mirror side, more than 0.3 m) and a 3-pole wiggler (drift next to the third or the fifth dipole, more than 0.35 m) as well. In addition, the lengths of quadrupoles are minimized, while keeping the required gradients well below their upper limits. Finally, a hybrid-7BA of 26.992 m is reached, with the layout and the optical functions (solid curves) presented in Fig. 1.

In the second step, longitudinal gradient is introduced into the outer dipoles and the emittance is further minimized. Each of the outer dipoles is split into five slices, which are considered to have different bending radii. Moreover, the bending angles are redistributed among the seven dipoles. With the other parameters unchanged, the analytical expression of the emittance is derived following Ref. [21] and is then minimized. As a result, the emittance is decreased from 100 pm·rad to 60 pm·rad. Since rectangular dipoles are used in the lattice and their lengths are unchanged, the C-S parameters remain the same. The variations in the bending radii of dipoles cause only a small change in dispersion functions, with the dispersion increasing slightly from 5.7 cm to 5.85 cm near the center of the dispersion bump (see the dashed curve in Fig. 1).

Three families of sextupoles (one family with horizontal focusing, SF, and the other two with vertical focusing, SD1 and SD2) are used for chromatic correction and are all located in the dispersion bump, where the relative high dispersion helps control the sextupole gradient and length to a reasonable level, i.e., below 8000 $T/m^2$ and ~0.3 m, respectively. Forty eight such hybrid-7BAs comprise the ring, with the main parameters listed in Table 1.

**3 Nonlinear optimization**

DA and MA are the two most important objectives of the optimization of nonlinear beam dynamics. The DA (for on-momentum particles) is most relevant to the injection efficiency, while the MA is most relevant to the Touschek lifetime, which is the main limitation of the available beam lifetime in a DLSR. Next we will show the study results of the nonlinear beam dynamics for this design.

As mentioned, by means of dispersion bump and local cancellation scheme, the chromatic sextupole strengths and the geometric aberrations induced by sextupoles have been controlled to a relatively small level. However, associated with the ultralow emittance and the small beta functions, it is still extremely difficult, if not impossible, to achieve DA of order of 10 mm to accommodate off-axis injection. Nevertheless, the requirement on DA can be greatly reduced with on-axis 'swap-out' injection scheme [22], in which the already-stored bunches are kicked out and replaced with fresh bunches from the booster. Based on above, the injection scheme for the ESRF-type design is chosen to be on-axis 'swap-out' injection in vertical plane by use of stripline kickers. Nevertheless, details about the injection system will be discussed elsewhere.

On the other hand, to save as much space as possible to accommodate kinds of hardware systems, except for three families of sextupoles, only one family of octupoles (with vertical focusing, OF) is used to correct the high-order aberrations, especially the vertical detune terms. It is known that at least two families of sextupoles with opposite focusing are needed to correct the natural chromaticity to a positive value (fixed to [0.5, 0.5] in this study). If the strength of one family of vertical-focusing sextupoles (e.g., SD2) is determined, it has unique solution for the other two families. There left only two free variables, and thus it is possible to make a global scan of all the possible settings in a two-dimensional variable space [e.g., the ($K_{SD2}$, $K_{OF}$) space] in a reasonable computing time.

Since the beam is considered to be injected in vertical plane, the goal is to maximize the vertical ring acceptance. To this end, the vertical DAs with all possible sets of ($K_{SD2}$, $K_{OF}$) are specifically calculated using the AT program [23], with the results shown in Fig. 2. Frequency map analysis [24] is also performed to analyze the stability of the motion. It reveals that since only the ideal lattice (without any error) is considered, the particle motion can cross the fatal first and second-order resonances without loss, which may cause an overestimation of the available ring acceptance for injected beam. To solve this problem, the vertical 'efficient' DAs that promises only the motions with fractional tunes within [0, 0.5] are calculated and presented in Fig. 2. It appears that the maximum vertical 'efficient' DA is about 4 mm and occurs around ($K_{SD2}$, $K_{OF}$) = (−150 $m^{-3}$, −5000 $m^{-4}$).

Similar scan is also performed for the MA, with the results shown in Fig. 3. In the calculation, it is assumed that the longitudinal dynamics is well optimized and the MA is limited by the transverse dynamics, or be more specific, the MA is determined by (the absolute value of) the momentum deviation that makes the fractional tune very close to or exactly at 0 or 0.5, in either *x* or *y* plane. One can see that the MA of 3% can be achieved around ($K_{SD2}$, $K_{OF}$) = (−100 $m^{-3}$, −1000 $m^{-4}$).

Fig. 4 shows the contour plots of both the vertical 'efficient' DA and the MA in the ($K_{SD2}$, $K_{OF}$) space. It seems impossible to find a solution that optimizes both two objectives simultaneously. Tradeoffs between the two objectives should be made. At last, a solution with ($K_{SD2}$, $K_{OF}$) = (−120 $m^{-3}$, −1600 $m^{-4}$) is chosen to provide a vertical 'efficient' DA of ~2.2 mm and a MA of ~2.4%. The on-momentum 'efficient' DA and the corresponding frequency map are shown in Fig. 5, and the off-momentum DAs are shown in Fig. 6. The coupling resonance $2\upsilon_x - 2\upsilon_y = 48\times3$ imposes strong perturbation on the motions in horizontal plane between [1.1, 1.4] mm, but has very weak impact on the motions in vertical plane. Needless to say, this solution is not an optimal result, but can basically meet the requirements of vertical on-axis injection and a moderate beam lifetime.

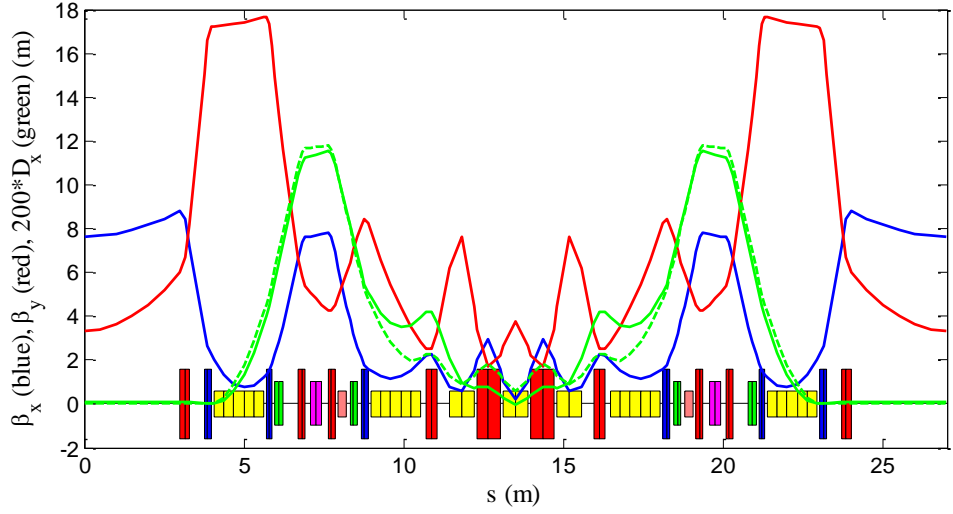

Fig. 1. (color online) Layout and optical functions of the hybrid-7BA designed for HEPS, without (solid curves) and with (dashed curves) longitudinal gradient combined in the outer dipoles.

Table 1. Main parameters of the ESRF-type lattice for HEPS

| Parameters | Values |
| --- | --- |
| Energy $E_0$ (GeV) | 6 (5) |
| Circumference $C$ (m) | 1295.616 |
| Horizontal damping partition number $J_x$ | 1.38 |
| Natural emittance $\varepsilon_0$ (pm.rad) | 60 (41.7) |
| Number of hybrid-7BA achromats | 48 |
| Maximum quadrupole gradient (T/m) | 80 |
| Maximum sextupole gradient (T/m$^2$) | 8000 |
| Number/length (m) of ID straight sections | 48/6 |
| Beta functions (m) in ID straight section (H/V) | 7.6/3.3 |
| Working point (H/V) | 113.20/41.28 |
| Natural chromaticity (H/V) | −149/−128 |
| Damping times (ms, x/y/z) | 18.8/26.0/16.0 |
| Energy spread $\sigma_\delta$ | $7.99\times10^{-4}$ |
| Momentum compaction $\alpha_p$ | $3.67\times10^{-5}$ |

Submitted to Chinese Physics C

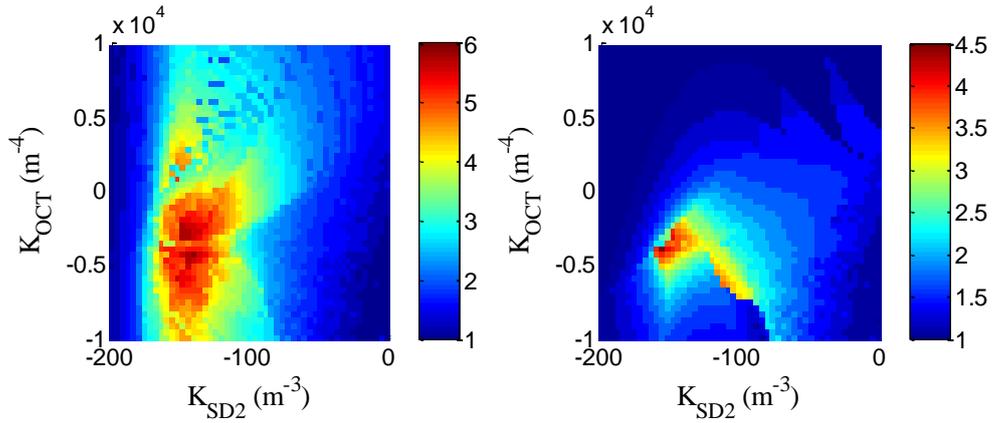

Fig. 2. (color online) Vertical DAs (left) and vertical 'efficient' DAs (right) with all possible sets of ($K_{SD2}$, $K_{OF}$). Different colors represent different DA sizes, in unit of mm.

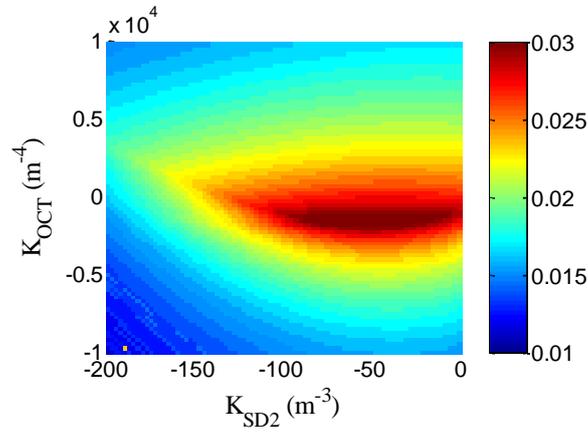

Fig. 3 (color online) MAs with all possible sets of ($K_{SD2}$, $K_{OF}$). Different colors represent different MA values, from 1% to 3%.

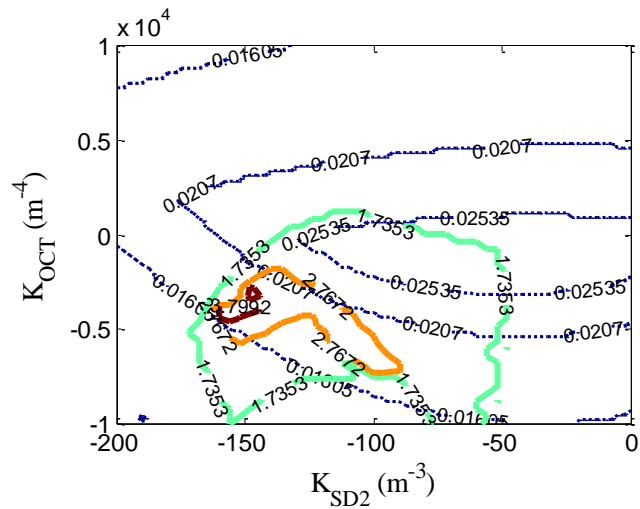

Fig. 4. Contour plots of the vertical 'efficient' DA (in unit of mm, solid curves) and MA (dotted curves) in the ($K_{SD2}$, $K_{OF}$) space.

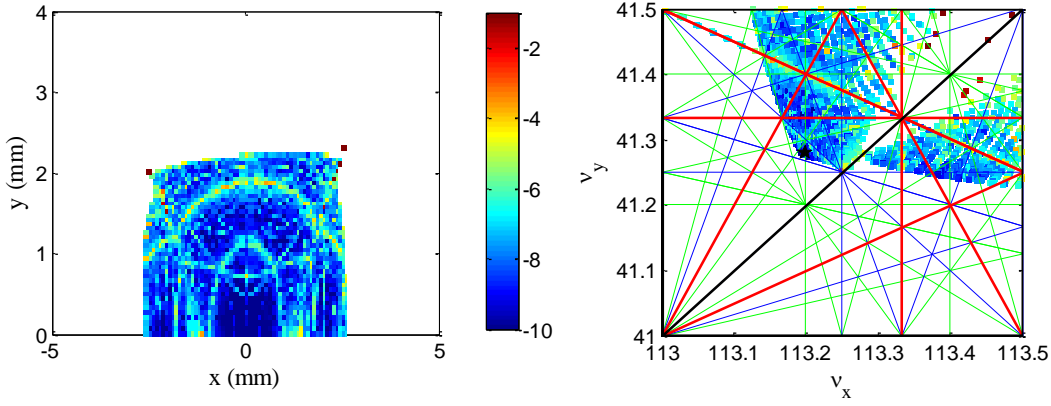

Fig. 5. (color online) The 'efficient' on-momentum DA and frequency map obtained after tracking over 1024 turns for the HEPS lattice design with ($K_{SD2}$, $K_{OF}$) = ($-120$ m$^{-3}$, $-1600$ m$^{-4}$). The colors, from blue to red, represent the stabilities of the particle motion, from stable to unstable.

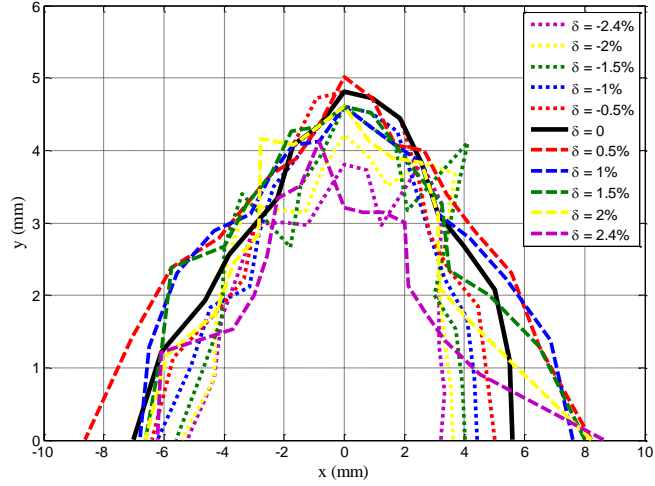

Fig. 6. (color online) Off-momentum DAs obtained after tracking over 1024 turns for the HEPS design with ($K_{SD2}$, $K_{OF}$) = ($-120$ m$^{-3}$, $-1600$ m$^{-4}$).

## 4 Conclusions

In this paper, we present an ESRF-type design for HEPS. The linear optics is designed such that the requirements of the ultralow low emittance, efficient chromatic correction, and enough drift spaces for various hardware systems are fully considered and basically satisfied. By using one family of octupoles to minimize the nonlinear driving terms caused by sextupoles, a large enough DA and moderate MA can be achieved for a promising ring performance.

We admit that there are still much room for further improvement of both the linear optics and the nonlinear beam dynamics. Nevertheless, the presented design can be served as a good baseline of HEPS for the future optimization and relevant studies.